\documentstyle[12pt,epsfig]{article}
\begin{document}

\title{
Underthreshold resonances in three-particle
molecular systems
}
\author{
F.M. Pen'kov\\
Joint Institute for Nuclear Research,\\
141980, Dubna, Russia \\
penkov@thsun1.jinr.ru}
\date{}
\maketitle

%PACS 03.65.Nk, 36.40+d

\begin{abstract}
To determine the lifetimes of Efimov states of negative two-atomic ions,
the problem of resonance scattering of a light particle on a pair of
identical particles has been considered. An analytic expression has been
obtained for resonance widths in the limit of forces of zero radius and low
binding energies in pairs. Calculations are compared with the numerical
solution of the Faddeev integral equations in a wide region of masses of
the light particle. It is shown that the widths of underthreshold
resonances in the scattering amplitude obtained from the integral equations
with the Yamaguchi  potential are well described by
the analytic expression, which
allows this expression to be used in the mass region inaccessible for
numerical calculations. It has been concluded that the lifetime of highly
excited negative molecular ions with a binding energy close to the
threshold of disintegration is practically unlimited.
\end{abstract}

\section{Introduction}
Recent papers \cite{Pena,Penb,Penc} have dealt with the
exotic states of systems of neutral atoms and an electron.
These states are characterized by large (tens of angstroms) sizes
and demonstrate the effects of three-body dynamics at low
binding energies in pairs. In particular, a direct analysis \cite{Pena}
of the Faddeev equations allowed us to determine the effective
potential of interaction of two neutral atoms in the presence of an
electron. This potential is local in the range\footnotemark
\footnotetext{\normalsize Here the Planck constant $ \hbar = 1 $  }
\begin{equation}
          max(r_0) \ll r \ll min( \kappa^{-1})
\label{region}
\end{equation}
($r_0$ is the range of two-body forces and $\kappa$ is the wave number
of a bound or virtual state of a pair) and contains both the long-range
components  of the type $ 1/r^2 $  typical for the Efimov effect
\cite{Efim1} and quasi-Coulomb terms of the type $1/r$ that essentially
contribute to the spectrum of negative two-atomic molecular ions with a
small binding energy (real or virtual) of an electron and an atom.
Specifically, experimental data on the scattering of an electron by a
helium atom allow us to conclude on a possible existence of a bound state
of the system He$_2^-$ even when there is no atom-atomic interaction
\cite{Penb}. Moreover, the analysis of a system of three neutral atoms of
alkali metals and an electron \cite{Penc} shows that the effective
interaction generated by the electron produce more than thousand bound
states. It has been assumed that similar systems can initiate clusters in
highly rarefied gases. However, that analysis is not complete for the atom-
atomic interaction with the binding energy higher than that of a
three-particle system. In this case, the system can decay into a molecule and
a free electron, and bound states of the three-particle system can
transform into resonances with the lifetime determined by widths of these
resonances.

The present paper is devoted to the study of these widths in the framework
of the three-particle problem of scattering of a light particle on a bound
pair of heavy particles. Like in refs. \cite{Pena,Penb},
we will consider the systems in which the lengths of two-body scattering are
much larger than the effective ranges of two-body forces satisfying the
conditions of the Efimov effect:
\begin{equation}
 \kappa r_0 \ll 1.
\label{uslov}
\end{equation}
The binding energy of atoms is taken finite; whereas that in electron-atom
pairs, zero for simplifying, analytic computations. The threshold of
rearrangement of the system coincides with the threshold of
disintegration and the above-mentioned spectrum of three-particle systems
\cite{Pena,Penb} is manifested as underthreshold Efimov resonances.
Resonances of that sort, but under the excitation threshold, in a three-boson
system were considered in ref.\cite{Pen1} and equidistance was shown both
for positions of resonances and their widths on the logarithmic scale.

In this paper, we consider the case when the mass of one particle is
considerably smaller than the masses of two other particles. Here we are
faced with almost classical motion of heavy particles
\cite{Pena}, which does not allow accurate numerical solution of the
Faddeev equations for the ratios of masses of an atom and an electron.
Therefore, like in ref.\cite{Pena}, numerical calculations are carried
out for the ratios of masses not exceeding 1/100 in order to verify
the analytic relation between the width and position of a resonance that
can be derived at extremely small effective ranges of two-body forces.
It is demonstrated below that the analytic expression satisfactorily
describes a highly complicated mass dependence of the resonance widths,
and thus it can be used to estimate the resonance widths for realistic
ratios of electron-atomic masses.

\section{Integral equations}

We consider a system of three spinless particles with masses
$m_i$, where the index $i=1,2,3$ denotes a particle, and with two-body
potentials $v_i$, whose index corresponds to a particle absent in a
two-particle subsystem. Identical particles 2 and 3 with masses $m_2 =m_3 =
m$ interact via the potential $v_1$ and form a bound state with the
energy $\varepsilon_1 = - \kappa_{1}^2/2m_{23}$, where $m_{ij}$ is the
reduced mass of particles $i$ and $j$.  Particle 1 with mass $m_1$
interact with particles 2 and 3 via the potentials $v_3$ and $v_2$ ($v_2 =v_3
= v$), respectively, which generate two-particle states (real or virtual)
with an extremely low binding energy ($\kappa_2 = \kappa_3 =\kappa
\to 0$). The potentials $v_i$ (between particles j and k) were taken to be
the separable Yamaguchi potentials acting only in the s-wave:
\begin{equation}
v_i(p,p') = - \frac{4 \pi}{m_{jk}}
\frac{\beta_i (\beta_i+\kappa_i)^2}{(\beta_i^2+p^2)(\beta_i^2+{p'}^2)}.
\label{Yam}
\end{equation}
The sign of the wave number $\kappa$ defines either a bound ($\kappa > 0 $)
or a virtual state of a pair. The parameter
$\beta_i$
determines the range of forces in a pair. $i$.
When
$\beta_i \gg \kappa_i$,
the expansion of the effective range in a pair
$i$ over the momentum of relative motion
$p$
($ p \ \mbox{cotan($ \delta_i $)} = \kappa_i + r_{eff} p^2/2 + ... $)
gives
$r_{eff}= 3/\beta_i $, which allows us to employ condition (\ref{uslov})
in the form $ \kappa_i \ll \beta_i $. In what follows for brevity, we use
the separable potential given by projectors:
$v_i = |\nu_i> <\nu_i|$.

Consider the scattering of particle 1 on a bound state of identical
particles 2 and 3 with the energy of relative motion
$ E_k = k^2_0/2 m_{1,23} $,
lower than the energy of rearrangement of the system
$ \varepsilon_2 - \varepsilon_1 $.
Here
$m_{i,jk}$ is the reduced mass of
particle i and a pair (j,k).
The momentum
$\bf k_i$
is always the momentum of relative motion of particle i and
pair (j,k), whereas the momentum
$ \bf p_i $
refers to motion inside the pair (j,k).
(If no confusion, the indices are omitted.)

To write the system of Faddeev integral equations
(see, e.g.,~\cite{Fad}), we need an expression for the product
of a two-body t-matrix in the 3-particle space and the free Green
function $G^0(Z) = (Z-h^0_{i}-h^0_{(i)})^{-1} $. The Hamiltonian of free
motion of three particles is given by two terms, the Hamiltonian of free
motion $h^0_{i}$ of particle i and pair (j,k) and the Hamiltonian of
relative motion in pair $h^0_{(i)}$. The quantity $ Z=E_k + \varepsilon_1 +i0
$ is the total energy of a 3-particle system. For separable potentials, this
product can be represented in the form \cite{Pen2} $$  t G^0(Z) =
v_i|\varphi_i> g^0_i(Z - \varepsilon_i) < \tilde \varphi_i |, $$
that includes the wave function of a two-particle bound state $| \varphi_i>$
and the function $ <\tilde \varphi_i(Z)| $ defined in the 3-particle space.
Hereafter, we make use of expressions for 2-particle Green functions
$g^0_i(x)$ and $g^0_{(i)}(x)$ corresponding to two-particle Hamiltonians
$h^0_{i}$ and $h^0_{(i)}$, respectively. In this notation, the projection of
the function $< \tilde \varphi_i |$ on $< {\bf k_i}|$ is of the simple form:
\begin{equation} < \tilde \varphi_i | = R_i <\varphi_i|v_i G^0
(Z-k_i^2/2m_{i,jk}); \label{phi} \end{equation} $$ R_i=-\frac{1}{2m_{jk}} \
\frac{(\beta_i+a_i)^2 (a_i+\kappa_i)} {(2 \beta_i +\kappa_i +a_i) <
\varphi_i|v_i| \varphi_i> }; \quad a_i= \sqrt{-2 m_{jk} (Z-k_i^2/2m_{i,jk})}.
$$ Note that the 2-particle $t$-matrix corresponding to a separable
potential of the Yamaguchi type has two poles. The near one is at the point
of a 2-particle bound state, $p_i = \pm i \kappa_i$ (on the physical sheet
for a realistic bound state and on the unphysical sheet for a virtual state).
The distant pole on the unphysical sheet $p_i = - i ( 2 \beta_i + \kappa_i$)
determines the range of two-body forces. The limit of zeroth range is
achieved with $\beta$ tending to infinity. In this limit, there remains one
pole, and therefore it does not matter, which two-body potential, either
local or nonlocal, has generated it. Here we use the separable potential
only for reasons of simplification of writing the Faddeev integral equations.
In this case, just $t$-matrices remain finite in the limit $\beta
\to \infty $.

Once the two-particle $t$-matrices are determined, we write the Faddeev
equations for the scattering of particle 1 on a bound pair of identical
particles 2 and 3 in the form
\begin{equation}
\begin{array}{c}
T^{el}  = V_{12} g^0_2(Z- \varepsilon_2) T^{r}, \\[3mm]
T^{r}  =2 V_{21} + 2 V_{21} g^0_1(Z- \varepsilon_1) T^{el} +
		      V_{23} g^0_3(Z- \varepsilon_3) T^{r},
\end{array}
\label{int_eq}
\end{equation}
where
$ V_{ij}=R_i^{1/2} < \varphi_i|v_i G_0(Z) v_j | \varphi_j>R_j^{1/2}  $,
and the transition matrices
$T^{el}$ and $T^{r}$ are connected with the physical amplitudes of
elastic and inelastic scattering by the simple formulae:
$$ f^{el} = -\frac{m_{1,23}}{2 \pi} \ T^{el}( {\bf k}_1^{out},{ \bf k}_1^{in}),
\qquad f^{r} = -\frac{\sqrt{m_{1,23} m_{2,13}}}{2 \pi} \
T^{r}( {\bf k}_2^{out},{ \bf k}_1^{in}),$$
where in- and out-momenta
$k^{in}$ and
$k^{out}$ are on energy surfaces.

The system of complex integral equations (\ref{int_eq}) can be transformed
into the system of equations real up to the threshold of rearrangement
(see, e.g., \cite{Bel}).
The simple change upon partial decomposition
\begin{equation}
f^{el}(k,k_0)=  -\frac{m_{1,23}}{2 \pi} \
K^{el}(k,k_0) \ \left(1 + i k_0 f^{el}(k_0,k_0) \right),
\label{amp}
\end{equation}
$$
f^{r}(k,k_0)= -\frac{\sqrt{m_{1,23} m_{2,13}}}{2 \pi} \
K^{r}(k,k_0) \ \left(1 + i k_0 f^{el}(k_0,k_0) \right)
$$
results in the system of equations for real functions
$K^{el}$ and $K^{r}$ solved numerically.
To distinguish the real equations with principal-value integrals, we
further denote the total energy as
$z= \mbox{Re} Z$.
In this notation, real integral equations for
$K^{el}$ and $K^{r}$
coincide with eqs. (\ref{int_eq}) when
$T^{el}$ is replaced by  $K^{el}$; $T^{r}$, by $K^{r}$; and $Z$, by $z$,
respectively:  \begin{equation} \begin{array}{c} K^{el}  = V_{12} g^0_2(z-
\varepsilon_2) K^{r}, \\[3mm] K^{r}  =2 V_{21} + 2 V_{21} g^0_1(z-
\varepsilon_1) K^{el} + V_{23} g^0_3(z- \varepsilon_3) K^{r}.  \end{array}
\label{int_rea}
\end{equation}
Just these equations for
$K^{el}$ and $K^{r}$
were solved numerically. Below, we show the scheme of origin of
underthreshold resonances on the basis of equations (\ref{int_rea}).

\section{Underthreshold resonances}

To see the origin of underthreshold resonances, we transform
eqs. (\ref{int_rea}) as follows:
\begin{equation}
\begin{array}{c}
K^{el}  = V_{12} g^0_2(z- \varepsilon_2) K^{r}, \\[3mm]
K^{r}  =2 V_{21} +  V g^0_3(z- \varepsilon_3) K^{r},
\end{array}
\label{int_rea2}
\end{equation}
where the effective potential of an (inelastic) channel closed in energy is
of the form
\begin{equation}
V= V_{23} + 2 V_{21} g^0_1(z- \varepsilon_1) V_{12}.
\label{pot}
\end{equation}
The system of equations  (\ref{int_rea2}) describes the two-channel process
of scattering of particle 1 on pair (2,3) vie one-channel interaction
of particle 2 (or 3) with a pair of particles 1 and 3 (or 2).
For this purpose, using system
(\ref{int_rea2}), we express $K^{el}$
in terms of the Green function of the closed (inelastic) channel
$ g_v(x) =(x-h^0_2 -V)^{-1} $:
\begin{equation}
   K^{el} = 2 V_{12} g_v(z-\varepsilon_2) V_{21},
\label{Kel}
\end{equation}
whose spectrum determines specific features of the elastic channel.
Resonances in elastic scattering correspond to points of the spectrum  $ E_t$
of the Hamiltonian
$ h_v=h^0_2 +V $
when
$ E_t > \varepsilon_1- \varepsilon_2 $. Though this statement is obvious,
we present the scheme of construction of the S-matrix for
resonances because of their peculiarities in the system under consideration.
To this end, we consider the case when the energy
$z$ is close to
$ \varepsilon_2 + E_t $ and separate the singular part
$$
g_v(z-\varepsilon_2)= \frac{|\Psi_t>< \Psi_t|}{\omega} + g_R,
$$
from the Green function $g_v$, where
$\omega= z-\varepsilon_2 -E_t $,
$ \Psi_t $ is the
wave function corresponding to the eigenvalue
$E_t$,
and $g_R$ is a regular part of the Green function usually neglected.
Then, using expression (\ref{Kel}) and connection of the physical
amplitude with $K^{el}$ (\ref{amp}), we derive the
$S$-matrix ($ S=1 + 2 i k_0 f^{el}(k_0,k_0)$) in the form
$$
S= \frac{1-iB}{1+iB} \
\frac{
\omega + \frac{1}{2} \frac{\Gamma B}{1+B^2} - i \frac{1}{2}\frac{\Gamma }{1+B^2}}
{\omega + \frac{1}{2}\frac{\Gamma B}{1+B^2} + i \frac{1}{2}\frac{\Gamma }{1+B^2} },
$$
where the width
\begin{equation}
\Gamma = 2 \frac{k_0 m_{1,23}}{\pi}| <k_0|V_{12}|\Psi_t> |^2 ,
\label{Gam}
\end{equation}
generated by the singular part of the Green function is changed by the
regular part: $B= \frac{k_0 m_{1,23}}{ \pi } <k_0|V_{12}g_R V_{21}|k_0> $.
The shift of a resonance from $E_t$ is also determined by that part. It is
just the quantity $B$ that determines the background scattering far off the
resonance when $S= \frac{1-iB}{1+iB}$. Introducing the background scattering
phase $ \delta_f = - \mbox{arctan(B)} $, we can rewrite the
$S$-matrix in the form
 \begin{equation} S= \exp{(2i \delta_f)} \frac{ \omega - i
\frac{\Gamma}{4} - \frac{\Gamma}{4} \left( \sin{(2 \delta_f)} +i \cos{(2
\delta_f)} \right) } { \omega + i \frac{\Gamma}{4} - \frac{\Gamma}{4} \left(
\sin{(2 \delta_f)} -i \cos{(2 \delta_f)} \right) }.  \label{Smat}
\end{equation}
It is to be noted that expression (\ref{Smat}) has been derived without
approximations and is a convenient representation of the $S$-matrix.
However, the Breit-Wigner parametrization, separation of the resonance
energy and width, imposes a certain constraint on widths.
They should be rather small so that an isolated resonance could be
considered. The true width of the resonance $\tilde \Gamma$ is
determined not only by $\Gamma$, but also by the phase of background
scattering: $ \tilde \Gamma = \Gamma \cos^2(\delta_f) $. Just for
this reason the expression for $\Gamma$ contains plane waves rather than the
wave functions of the state scattering in a certain background potential.
Below, we consider an expression for $\Gamma$ that is always larger than or
equal to  $ \tilde \Gamma$. Other specific properties of the $S$-matrix
represented by (\ref{Smat}) are rather obvious and their discussion goes
beyond the scope of this paper.

\section{Wave function of the closed channel}
We are interested in a series of resonances corresponding to the Efimov
effect, i.e. when conditions (\ref{uslov}) hold valid and the spectrum
of the Hamiltonian
$h_v$
condenses around zero. In this case
the potential $V$ gets much simplified, and for
studying underthreshold resonances it is sufficient
to consider the $S$-part of the effective potential at
low energies, or more precisely, at momenta
$k \gg \sqrt{(-2 m_{2,13}z)}$,
where solutions to the Schr\"odinger equations slightly depend on energy.
Under this auxiliary condition
($\beta \to \infty, \ z \to 0 $)
the terms of the effective potential
(\ref{pot}), the "exchange" potential
$ V_{ex} \equiv V_{23}$
and two "triangular" (with internal integration) potentials
$ V_{tr} \equiv V_{21} g_{1}^0(z- \varepsilon_1) V_{21}$,
can be written in the form:
\begin{equation}
V_{ex}^0(k,k') = - \frac{\pi}{2 \sqrt{kk'} \lambda_1} \frac{1}{\sqrt{m_{3,12}m_{12}}}
\ln \left(\frac{k^2+k'^2 + 2 \lambda_1 k k'}{k^2+k'^2 - 2 \lambda_1 k k'} \right),
\label{ex}
\end{equation}
\begin{equation}
V^0_{tr}(k,k') = - \frac{1}{\sqrt{kk'} (2 \lambda_1)^2} \ \frac{m_{12}}{m_{23}^2} \
\mbox{v.p.}\int\limits^{\infty}_0 \frac{ {\rm}d t}{t-a} L(t,k) L(t,k');
\label{tr}
\end{equation}
$$
L(t,k) = \ln \left(\frac{\gamma t^2+k^2 + 2 \lambda_2 k t}{\gamma t^2+k^2 - 2 \lambda_2 k t} \right),
\
a=\kappa_1 \sqrt{\frac{m_{1,23}}{m_{32}}}, \
\lambda_i =\sqrt{\frac{m_{ij} m_{ik}}{m_i^2}}, \
\gamma= \sqrt{\frac{m_{23}}{m_{12}}}.
$$
The upper index 0 means that the effective interaction
is considered at $z=0$.
The exchange potential
$V^0_{ex}$ describes the
scattering of particle 2 (or  3) on a bound pair of particles
1 and 3 (or 2) at the zeroth binding energy in the pair and
is a ''classical'' potential generating Efimov states.
Solutions in the field of that potential investigated for a
three-boson system \cite{Pen2},
coincide with solutions in the field of
a local potential of the type
$- (\mu^2 + 0.25)/2m_{2,13} \rho^2$  ($\rho$~is the Jacobi coordinate
of relative motion of a particle and a pair) with the coupling constant
$\mu$ obeying a transcendental equation. Below, we derive an equation of
that sort for the total interaction potential. Note that the
potential
$V^0_{ex}$ behaves like $1/k$ and admits a solution in the form
$k^{s}$.
The potential
$V^0_{tr}$
is more complicated in form and admits solutions like that
only in the region of momenta where the quantity
$a$ in the denominator of the integrand
of expression (\ref{tr}) can be neglected.
To verify this, we look for solutions to the Schr\"odinger equation
with the zeroth energy in the form
$\Psi_t = k^{i \mu - 5/2}$. To this end, we define the quantity $\Pi$:
$$
\Pi(k) = - \frac{2 m_{2,13}}{k^{i\mu-1/2}} \
\int V^0(k,t) t^{i\mu-5/2} \frac{ d^3 t}{(2 \pi)^3},
$$
and write it with an index corresponding to the potential.
This quantity represents the ratio of contributions of the
potential and kinetic energies to the Schr\"odinger equation.
Then, the Schr\"odinger equation is written in the form:
$
\Pi_{ex} + 2 \Pi_{tr} =1.
$
The contribution of the exchange potential is expressed via the integral
$$ \Pi_{ex} = \frac{1 + \zeta_1^2}{\zeta_1} \ I_1, \qquad
 I_1=\frac{1}{2 \pi} \int\limits^{\infty}_0 x^{i \mu-1}
\ln \left(\frac{1+2 \lambda_1 x + x^2}{1-2 \lambda_1 x + x^2}\right) d x,
$$
($ \zeta_i = \sqrt{\frac{m_j m_k}{m_i M}} $, $M$ is the total mass)
that exists at
$ -1 < \mbox{Im} \mu < 1 $ and is taken over residues upon integrating
by parts:
$$
I_1   = \frac{\mbox{sh}(\mu \ \mbox{arctan} \zeta_1)} {\mu \ \mbox{ch}(\frac{\pi}{2}\mu)}.
$$
Introducing the function $\Phi$ in the form
$$
\Phi_i(\mu) = \frac{1 + \zeta_i^2}{\zeta_i}
\frac{\mbox{sh}(\mu \ \mbox{arctan} \zeta_i)}
{\mu \ \mbox{ch}(\frac{\pi}{2}\mu)},
$$
we obtain
$$
\Pi_{ex} =  \Phi_1(\mu).
$$

The function $\Pi_{tr}(k)$ is, upon simple transformations,
expressed in terms of the function $\Phi$ and a new
function $\tilde \Phi$:
$$
\Pi_{tr}(k) =  \Phi_2(\mu) \ \tilde \Phi,
$$
$$
\tilde \Phi = \frac{1 + \zeta_2^2}{2 \pi \zeta_2} \ \mbox{v.p.}
\int\limits^{\infty}_0 x^{i \mu-1} \frac{k x}{k x -a\gamma}
\ln \left(\frac{1+2 \lambda_2 x + x^2}{1-2 \lambda_2 x + x^2}\right) d x.
$$
This integral is easily transformed into contour
integrals along two logarithmic cuts and integration contours
can always be chosen so that either
$|x| \geq 1 $ or $|x| \leq 1 $. This allows decomposition of the
denominator of the integrand in a series either at large
($k > a \gamma $) or at small momenta ($k < a \gamma $ ).
In the first case, we obtain the leading term $\tilde \Phi = \Phi_2$;
whereas in the second, $\tilde \Phi \propto k/\kappa_1$.
Therefore, we can write the transcendental equations for
$\mu$ in two asymptotic regions. Considering that
$ \Pi_{ex} + 2 \Pi_{tr} = 1$,
we get
\begin{eqnarray}
 \Phi_1(\mu) =1, & k \ll \sqrt{\frac{m_{1,23}}{m_{12}}} \kappa_1 ;
\label{inf}
\end{eqnarray}
\begin{eqnarray}
 \Phi_1(\mu) + 2 \Phi_2^2(\mu) =1, &
k \gg \sqrt{\frac{m_{1,23}}{m_{12}}} \kappa_1.
\label{zero}
\end{eqnarray}
Owing to the function $\Phi$ being even in $\mu$, the
wave function $\Psi_t$ is a linear combination of
$k^{\pm i\mu -5/2}$,
thus coinciding with the Fourier transform of the wave
function in the field of the local potential
\begin{equation}
V_{ef}(\rho) = - \frac{\mu^2 +0.25}{2 m_{2,13} \rho^2},
\label{Veff}
\end{equation}
when $|V| \gg |z|$, with different coupling constants
at small and large $ \rho $.
Solutions in the field of $1/\rho^2$-type potential are well known.
Specifically, the energy
levels obey the relation
$ E_{n-1} /E_{n} = \eta $, where
the quantity $\eta$ depends merely on $\mu$:
\begin{equation}
 \eta = \exp(2\pi/\mu),
\label{eta}
\end{equation}
and they either go to minus infinity (fall onto the center indicated by
Thomas \cite{Thom}, or are condensed to zero from below
(the Efimov effect \cite{Efim1}). In our case, deep levels
correspond to small distances and are determined by the
value of $\mu_2 $ that obeys the equation (\ref{zero}). Shallow levels
with $ z > \varepsilon_1 $ in the scattering channel produce resonances
and are determined by the value of  $\mu_1$ from equation (\ref{inf}).

A simple analysis of eqs.(\ref{inf}) and  (\ref{zero}) shows that
$\mu_2$ exponentially tends to $\mu_1$ when $m_1/m_2 \to 0$, and
the quantity $\mu_1$ in this limit is described by the expansion:
\begin{equation}
\mu_{as} =  c \zeta_1 + \frac{c}{(c+1)\zeta_1} +
O \left( \frac{1}{\zeta_1^3} \right)
\label{mu_as}
\end{equation}
($\zeta_1=\sqrt{m_2 m_3/m_1 M} \gg 1 $),
coinciding with the limit found earlier \cite{Pena}.
The constant $c=0.5671...$ obeys the equation
$c=\exp(-c)$.
We do not present detailed computations and cite the dependences
of   quantities $\mu$
on the ratio of masses of a light and a heavy particle in Table \ref{tab1}.
Due to the rapid convergence of $\mu_2 $ to $\mu_1$, the second
term in eq. (\ref{zero}) gets negligible (solutions differ in the sixth
digit at $m_2/m_1=100$), and thus, in the limit $m_1/m_2 \ll 1$ the second
term in the potential $V$ can be neglected. To demonstrate the smallness,
we present the ratio of potential energies $\Pi_{ex}(\mu_1)$ and
$\Pi_{tr}(\mu_1)$.
%%%%%%%%%%%%%%%%%%%%%%%%%TAB 1 %%%%%%%%%%%%%%%%%%%%%%%%%%%%%%%%%
\begin{table}[h]
\caption{
The dependences of $\mu_1$, $\mu_2$, $\mu_{as}$ on masses (for explanation,
see the text)
}
\vspace{5mm}
\begin{tabular}{|c|c|c|c|c|c|c|}
\hline
$m_2/m_1$  & 10       & 20       & 30       & 50       & 70       &100\\
\hline
$\mu_1$     & 1.379051 & 1.893909 & 2.284713 & 2.906777 & 3.415826 & 4.061110\\
$\mu_2$     & 1.468174 & 1.919444 & 2.293865 & 2.908434 & 3.416219 & 4.061172\\
$\mu_{as}$  & 1.430016 & 1.907906 & 2.289978 & 2.908096 & 3.416437 & 4.061489\\
$ \Pi_{tr}/\Pi_{ex}$
	    & 1.12 10$^{-1}$
		       & 2.20 10$^{-2}$
				  & 6.41 10$^{-3}$
					     & 9.01 10$^{-4}$
							 & 1.81 10$^{-4}$
                                                          	   & 2.38 10$^{-5}$ \\
\hline
\end{tabular}
\label{tab1}
\end{table}
%%%%%%%%%%%%%%%%%%%%%%%%%%%END TAB1 %%%%%%%%%%%%%%%%%%%%%%%%%%%%%%%%

Thus, the wave function $\Psi_t$ for a very light particle 1
in a large region
($k \gg \sqrt{(-2 m_{2,13}z)}$) of momenta is a linear combination
of the functions
$k^{\pm i\mu -5/2}$ with the quantity
$\mu=\mu_1$. This momentum asymptotics is sufficient for
computing the matrix elements in the width $\Gamma$ given by
(\ref{Gam}) since a dominant contribution to these integrals comes from the
region of momenta $\sim \kappa_1$. The difficulty in normalizing  the
wave function can be overcome in a simple manner. Our wave function
describes the motion in the effective potential (\ref{Veff}). Solutions  in
the field of that potential are well known: $\Psi_t(\rho) \sim
K_{i\mu}/\sqrt{\rho}$, they are expressed in terms of modified Bessel
functions. The function $\Psi_t$ can be normalized, its Fourier transform
and coefficients of the leading asymptotic terms $k^{\pm i\mu -5/2}$
can be obtained. This scheme is somewhat cumbersome but produces quite a
reasonable result that for a small mass of particle 1 coincides with
the exact result. The latter (at zeroth two-body forces) can be obtained
more easily: the solution to the integral equation with potential
$V_{ex}$ can be derived in the whole region of momenta, without
setting $z=0$, by the Mellin transformation, like it has
been carried out in refs.
\cite{Dan1,Fad1} in studying the properties of the
Skornyakov-Ter-Martirosyan equation  \cite{STM}.
In ref. \cite{Fad1}, an approximate wave function has been found for a system
of three bosons with zeroth two-body forces which at zeroth binding
energies in subsystems becomes exact. The Lippmann--Schwinger equation
with potential $V_{ex}$ throughout the whole momentum region for the
function $\psi = k(k^2-2m_{2,13}z)^{3/4}\Psi_t $, upon making
$t=k/\kappa_t$ ($\kappa_t=\sqrt{-2m_{2,13}z}$) to be dimensionless,
differs from the Skornyakov--Ter-Martirosyan for three bosons by factor
2 and different values of the mass constants $\lambda_1$ and
$\alpha=\sqrt{m_{13}/m_{2,13}}$ :
\begin{equation}
\psi(t)=\frac{1}{2 \pi \lambda_1 \alpha}
\int \limits^{\infty}_0 dt' \ln
\left (
\frac{t^2 +t'^2 +2 \lambda_1 t t' + \alpha^2}
{t^2 +t'^2 - 2 \lambda_1 t t' + \alpha^2}
\right )
\frac{1}{\sqrt{t'^2+1}} \psi(t').
\label{LSh}
\end{equation}
Therefore, equation (\ref{LSh}) has also an analytic equation that
can be derived by the Mellin transformation. Here we present the
final result:
$$
\psi(t) = A \sin(\mu \ln(t+\sqrt{1+t^2})),
$$
that can be verified by a simple substitution of
$\psi$ into eq. (\ref{LSh}). The constant $\mu$
again obeys equation (\ref{inf}), and the constant
$A$ is defined by normalization. The integral equation
(\ref{LSh}) is solvable at all energies, however, the spectrum
can be fixed by imposing constraints on the coefficients
of  $\sin(\mu \ln k) $ and
$ \cos(\mu \ln k)$ at large $k$ \cite{Dan1,Fad1}:
$$\sin(\mu\ln\kappa_t) = b \cos(\mu\ln\kappa_t).$$
An arbitrary constant $b$ determines the wave number
$\kappa_t^{(n)}$ for a level with number $n$:
$$\kappa_t^{(n)} = \exp(\pi n/\mu + \mbox{arctan(b)}/\mu).$$
The spectrum thus found satisfies equation
(\ref{eta}) and demonstrates the effect of fall on the center at
large positive $n$, noted in ref. \cite{Fad1}, and the Efimov effect
(logarithmic condensation of levels around $z=0$) at large negative
$n$, later discovered by a different method \cite{Efim1}.

Normalizing the wave function $\Psi_t$ over the whole momentum space,
we arrive at the final result for the wave function of the closed
channel:
\begin{equation}
\Psi_t(k)=\frac{2\pi}{
\sqrt{
\kappa_t^3 (1- \pi \mu / \mbox{sh}( \pi \mu))
}
}
\quad
\frac{1}{t(1+t^2)^{3/4}} \sin(\mu \ln(t+ \sqrt{1+t^2})).
\label{wf}
\end{equation}

\section{Widths of underthreshold resonances}
As mentioned above, the width of a resonance $\tilde \Gamma$ depends on
the phase of background scattering and on $\Gamma$ (\ref{Gam})
that is defined via matrix elements of the potential
$V_{12}$ that connects the open and closed channels of elastic
scattering. The background phase depends on details of short-range
two-body forces and cannot be considered (has no limit) at the zeroth
ranges of the forces, whereas the quantity
$\Gamma$, being an upper limit of the resonance width, can in
this limit
($\beta \to \infty$) be calculated. In view of the equality
$E_k+ \varepsilon_1 = E_r$ for low
$|E_r|$ (as compared to $|\varepsilon_1|$) we can set
$k_0 \to \kappa_1 \sqrt{m_{1,23}/m_{23}}$. Owing to a simple relation
$E_r=E_t + \Gamma \sin(2 \delta_f)/4$
(see (\ref{Smat})) and small widths $\Gamma$, in the expressions
given below we put $E_r=E_t$.
Then considering the definition of $V_{12}$ and the width given
by (\ref{Gam}), trivial computations result in the expression
\begin{equation}
\frac{\Gamma}{|E_t|} = 32 \pi
\left(
\frac{m_{2,13}}{m_{23}}
\right)^2
\left | J \right |^2,
\label{GE1}
\end{equation}
where $J$ is given by the integral over the whole momentum
space:
$$
J=
\frac{\kappa_1}{\kappa_t}
\int \frac{d^3 k}{ (2 \pi)^3 }
\frac{(k^2+\kappa_t^2)^{1/4}}
{
\kappa_1^2 + k^2 - {\bf k_0 k} + \frac{1}{4} k_0^2
}
\Psi_t(k).
$$
By using the wave function of the closed channel (\ref{wf}),
we can compute this integral analytically. Since this procedure
is quite complicated, we here only outline it. After a trivial
angular integration and making  $t=k/\kappa_t$ dimensionless, we should
make the change of variables $t=(x^2-1)/(2x)$ that allows us to get
rid of radicals and transforms the integration range from
(0, $\infty$) to (1, $\infty$). The integrand is invariant under
the transformation $x \to 1/x$, which permits a back change to the
interval (0, $\infty$). Upon integration by parts we get rid of the
logarithmic function, and the integral~$I$
$$
 I=\int \limits_0^{\infty} x^{i\mu} \frac{Q_1(x)}{Q_2(x)} dx,
$$
where $Q_1$ and $Q_2$ are polynomials providing convergence at zero and
infinity, is computed in a standard way, the integral over the upper
edge of the power cut is expressed via a contour integral along the cut.
The latter can be closed at infinity and expressed in terms of residues
at zeros of the polynomial $Q_2$.

Upon that integration, we obtain the following expression for
$J$:
\begin{equation}
J=\frac{1}{2 \sqrt{1-\pi \mu/ \mbox{sh}(\pi \mu)}}
\frac{\mbox{sh}(\mu \ \mbox{arctan}(\zeta_2))}{\mu \zeta_2}
\frac{\sin(\mu \ln(2 \sqrt{\varepsilon_1/ E_t}))}
{\mbox{ch}(\pi \mu/2)}.
\label{J}
\end{equation}
Note remarkable properties of expressions (\ref{GE1}) and (\ref{J}). First,
the relative width $\Gamma/|E_t|$ does not depend on the number of a
resonance as positions of resonances are connected by expression (\ref{eta}),
 as a result of which for resonances with  numbers $n_{i}$ and $n_{i+k}$
the quantity $J$ differs by a factor $(-1)^{k}$. Therefore, on the
logarithmic scale, not only the positions of resonances but also their
widths are equidistant.  Second, the relative width of resonances
exponentially decreases with the mass of particle 1. To verify this, we
write a limiting expression
($m_1/m_2 \ll 1$) for the relative width. Considering
(\ref{mu_as}) and that
$$
\left(
\frac{\mbox{sh}(\mu \ \mbox{arctan}(\zeta_2))}{\mu \zeta_2}
\right)^2 \to 1.024... ,
$$
that limit can be represented in the form
\begin{equation}
\frac{\Gamma}{|E_t|} \simeq 32 \pi 1.024 \
 \exp
\left(- \pi c \sqrt{\frac{m_2}{2m_1}  }
\right)
\
\sin^2
\left(
\frac{\mu}{2}
\left(
\ln 4
\frac{\varepsilon_1}{E_t}
\right)
\right).
\label{GElim}
\end{equation}

As the observed width $\tilde \Gamma$ is always smaller than $\Gamma$,
we can make an upper limit on the width of a threshold resonance
\begin{equation}
\tilde \Gamma \ < \          103.0 \  {|E_r|}   \
 \exp
\left(- 1.260 \sqrt{\frac{m_2}{m_1}  }
\right),
\label{Gelim2}
\end{equation}
where we go back to the notation $E_r$ since the difference between $E_t$ and
$E_r$ is exponentially small. With expression (\ref{Gelim2}), we can
estimate lifetimes of strongly excited negative
ions of two-atomic
molecules. Already for the proton--electron mass ratio and certainly
overestimated scale of the binding energy of subsystem ($|E_r|$) in 1 eV,
the lifetime of a resonance ($1/\tilde \Gamma $) is larger than $10^{6} \ c $
and exceeds all possible relaxation times in gases. More interesting are
systems of a low energy of affinity of an electron and an atom. As mentioned
in the Introduction, these systems can be negative ions of molecules of
alkali metals. Inserting the mass ratio for lithium into expression
(\ref{Gelim2}) we obtain the lifetime $10^{40}\ c$ much greater than that
of the Universe.

\section{Numerical solutions}

To demonstrate the validity of analytic computations, we numerically solve
the system of Faddeev integral equations
(\ref{int_rea}) describing the scattering of particle 1 on a pair of
identical particles (2 and 3). Computations were performed by the
scheme described earlier \cite{Pen1} in which scattering can be considered
as close as possible to the reaction threshold.
%%%%%%%%%%%%%%%%%%%%%%%%%% Fig. 1 %%%%%%%%%%%%%%%%%%%%%%%%%%%%
\begin{figure}[h]
\begin{center}
\mbox{\epsfig{file=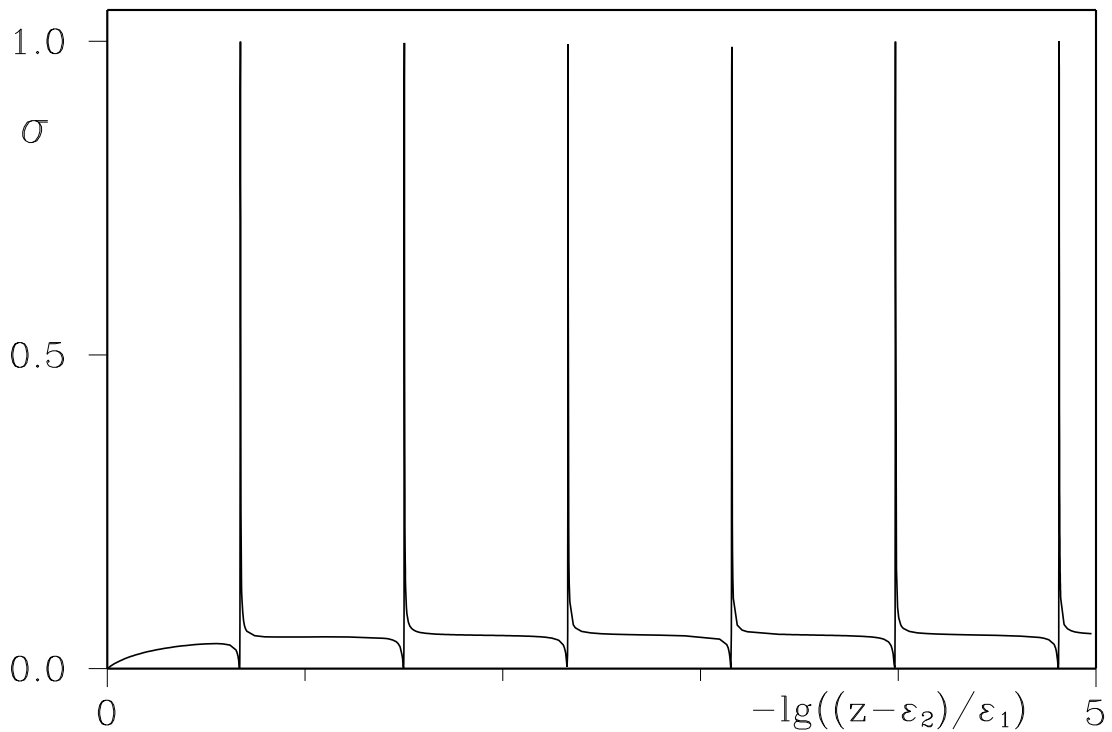,width=15cm}}
\caption{
Scattering cross sections for $m_2/m_1=65$. Explanations are given in
the text
}
\end{center}
\end{figure}
%%%%%%%%%%%%%%%%%%%%%%%%%End Fig 1 %%%%%%%%%%%%%%%%%%%%%%%%%
Our calculations do
not require such maximal proximity since the coupling
$\mu$ is large and we restrict our consideration to the region from zero
kinetic energy
($ z=\varepsilon_1 $) up to an energy being remote from the rearrangement
threshold by $10^{-8}$ MeV ($z-\varepsilon_2=-10^{-8}$ MeV). The
rearrangement threshold differs from the three-particle one by
$10^{-15}$ MeV ($ \varepsilon_2=-10^{-15} $ MeV), and the binding energy
in pair (2,3) $\varepsilon_1$  equals $-10^{-3}$ MeV.
The masses of particles 2 and 3 are equal to the nucleon mass. The constant
$\beta$ that determines the interaction range is taken the same for all
two-body potentials and equals 0.72  fm$^{-1}$. Measurement units for
the energy (MeV) and distances (fm) are taken in accord with the
Yamaguchi potential with parameters of the nuclear problem (masses and
interaction ranges) and characteristic potential energy in several ten
MeV. So, numerical calculations were based on small but finite interaction
ranges of two-body forces and binding energies in pairs. As the
problem is completely specified by the ratio of dimensional quantities, the
results of calculations are also valid for atomic units. For instance,
conditions
(\ref{uslov}) for pairs $$ \kappa_1 / \beta_1 \simeq 6.7 \cdot
10^{-3}; \qquad  \kappa_2 / \beta_2 = 10^{-6} \sqrt{m_1/m_2} \ \kappa_1 /
\beta_1 , $$ on the atomic scale of interaction ranges of two-body forces in
1 \AA \ \ ($r_0 \simeq 3/\beta $) correspond to the length of atom-atomic
 scattering ($1/\kappa_1 $) in about  50  \AA.

In Fig.1, we plot the scattering cross sections normalized to the  unitary
limit in S-wave, $\sigma_{un}=4 \pi /k_0^2 $, for the mass ratio
$m_2/m_1=65$. Since resonances condense to z=0, the energy is plotted on
the logarithmic scale. From Fig. 1 one can clearly observe the equidistance
of resonance positions. Unfortunately, even the maximum width in this mass
region of particle 1 (see Fig.2) remains very small and resonances are
practically straight lines.

In Table \ref{tab2} we report positions of the first
six resonances, the ratio of the energy of a preceding resonance to that
with a given number, i.e., the quantity $\eta_{calc}$,
and relative widths of resonances. For comparison, we present the values
of $\mu_1$ and $\mu_2 $, solutions to equations
(\ref{inf}) and (\ref{zero}), respectively: $\mu_1=3.29589...$,
$\mu_2=3.29654...$. They correspond to the values of $\eta$ given by
(\ref{eta})):
$\eta(\mu_1)=6.72823...$  and $\eta(\mu_2) = 6.72608...$.
%%%%%%%%%%%%%%%%%%%%%%% TAB 2 %%%%%%%%%%%%%%%%%%%%%%%%%%%%%%%%%%%%%%%
\begin{table}[h]
\caption{
Parameters of the first six resonances at $m_2/m_1=65$.
Explanations are given in the text
}
\vspace{5mm}
\begin{center}
\begin{tabular}{|c|c|c|c|c|}
\hline
$ N $   & $ E_r/\varepsilon_1 $  & $ \eta_{calc}$ & $\Gamma/|E_r|$ \\
\hline
1       & 2.1236 10$^{-1}$         &         & 4.413 10$^{-3}$ \\
2       & 3.1480 10$^{-2}$         & 6.746   & 4.246 10$^{-3}$ \\
3       & 4.6787 10$^{-3}$         & 6.728   & 4.208 10$^{-3}$ \\
4       & 6.9540 10$^{-4}$         & 6.728   & 4.202 10$^{-3}$ \\
5       & 1.0338 10$^{-4}$         & 6.727   & 4.201 10$^{-3}$ \\
6       & 1.5369 10$^{-5}$         & 6.726   & 4.200 10$^{-3}$ \\
\hline
\end{tabular}
\end{center}
\label{tab2}
\end{table}
%%%%%%%%%%%%%%%%%%% END TAB2
From the Table it is seen tat the energy ratio of the second
to the third resonance is in the region of analytic values
of $\eta$ up the fourth decimal point. Note that the first
resonance is quite far from the threshold. The last column demonstrates
the relative widths to be independent of the number of a
resonance.  Starting from the third resonance, changes occur in the
fourth decimal place. So, Figure 1 and Table 2 confirm the logarithmic
equidistance of the positions of resonances and their widths that
manifests itself in the constant character of relative widths of
resonances.

%%%%%%%%%%%%%%%%%%%%%%% Fig. 2 %%%%%%%%%%%%%%%%%%%%%%%%%%%%%%%%
\begin{figure}[h]
\begin{center}
\mbox{\epsfig{file=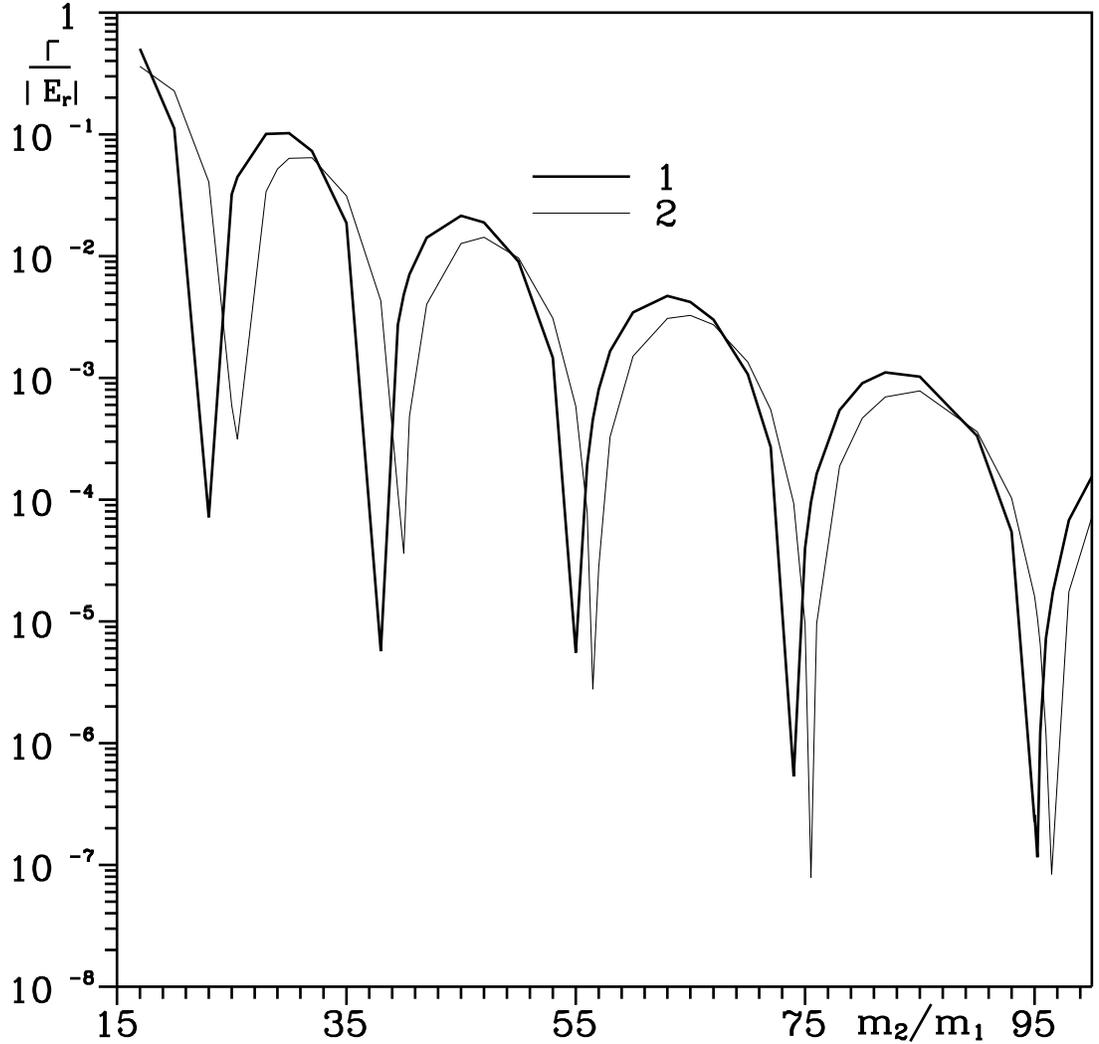,width=15cm}}
\end{center}
\caption{
Relative widths of resonances. 1 is a numerical solution
of the Faddeev equations; 2, the analytic dependence. Explanations are
given in the text.
}
\end{figure}
%%%%%%%%%%%%%%%%%%%%%%%%%End Fig. 2 %%%%%%%%%%%%%%%%%%%%%%%%%

To demonstrate the validity of expressions
(\ref{GE1}) and (\ref{J}) in describing the analytic
connection of widths of resonances and their positions at small ratios of
masses $m_1/m_2$, in Fig.2 we draw the relative widths obtained from
numerical solutions to integral equations (a solid curve)  and from
expression (\ref{GE1}) by substituting numerically found values of
resonance energies into it (a thin curve).
Analytic dependence of the resonance widths on masses follows the
curve though it has a small phase shift. Note
some peculiarities of the dependence of widths on the mass of
a light particle 1. First, the width of a resonance is not a
monotone function of its mass. Second, the maximum of the widths
of resonances exponentially decrease with the mass of particle 1
in accordance with the estimation (\ref{GElim}). In particular, for the range
of mass changes ($m_2/m_1$), from 15 to 100,  shown in the Figure, the
widths of resonances fall by three decimal orders. A strong nonmonotone
character of relative allows us to see the difference of analytic¨ ®â
estimations from the numerical results. If we estimate on the basis of sharp
local minima, then the observed phase shift for the minimum amount to
2 mass units and monotonically decreases to 1.25 units in the region
of the last minimum in the Figure ($m_2/m_1 \sim 95 $). Thus, the difference
between analytic and numerical estimates diminishes with mass
$m_1$.

\section{Conclusion}
So, we have estimated lifetimes of the Efimov
states of negative ions on the basis of the scattering problem of a light
particle (an electron) on a bound pair of two heavy particles (molecules)
under condition (\ref{uslov}). This three-body problem allows one to derive
analytic expressions for widths and positions of resonances and at the
same time to verify them by a direct numerical solution of the Faddeev
equations. For the first time, we have computed Efimov resonances under
the threshold of rearrangement of a three-particle system.

Our consideration leads us to conclusion that lifetimes of a new class of
molecular states discussed in refs. \cite{Pena,Penb,Penc} testify in
favour of them being bound states for all types of physical processes.
The work was performed within in project ISTC K-40.

\end{document}